\documentclass{kluwer}    

\begin{document}                
\begin{opening}         
\title{A Large Scale Energy Source For Feeding\newline{ISM Turbulence in Spiral
        Galaxies}}
\author{Xiaolei \surname{Zhang}}  
\runningauthor{Xiaolei Zhang}
\runningtitle{A Large Scale Energy Source for Turbulence}
\institute{SSAI/NASA's Goddard Space Flight Center
\newline Current Affiliation: Naval Research Lab}
\date{}

\begin{abstract}
One of the important consequences of a newly discovered
secular dynamical evolution process of spiral galaxies
(Zhang 1996,1998,1999) is that the orbiting disk matter receives energy
injection each time it crosses the spiral density wave crest.  
This energy injection  has been shown to be able to quantitatively
explain the observed age-velocity-dispersion relation of the
solar neighborhood stars. We demostrate in this paper that 
similar energy injection into the interstellar medium could 
serve as the large-scale energy source to continuously power 
the observed interstellar turbulence and to offset its downward
cascade tendency.  
\end{abstract}
\end{opening}         

\section{Introduction}
It has been several decades since it was first noticed
that Galactic molecular clouds and complexes
appear to be in a well-organized hierarchy, with their sizes
and velocity dispersions following a power-law
correlation , for cloud sizes ranging from 1 kpc which is the size of
the Giant Molecular Cloud (GMC) and HI Complexes, down to 0.1 parsec 
which is the size of the cores of low mass star-forming regions.
This observed hierarchy of cloud sizes and velocity dispersons
has been speculated to be produced by a
hierarchy of turbulent energy cascade (Larson 1981). 

Due to the natural tendency of turbulence to cascade downward
from large to small scales, its maintenance
requires continuous kinetic energy injection at larger scales.
Many candidate mechanisms for energy injection into the interstellar
medium (ISM) have been envisioned over the
past few decades.  It is generally agreed that the relevant
mechanism has to at least provide some means of 
energy injection from the largest scales, i.e. $\sim $ 1 kpc. 
Small-scale energy injection mechanisms alone
generally have difficulty in producing the observed large-scale correlation;
the resultant hierarchy also has velocity dispersions
largest on the smallest scales and smallest on the largest scales,
contrary to the observed trend.  

The most obvious reservoir of turbulent energy on the largest-scale 
is of course the galactic rotation.  
One problem with the past-proposed means of tapping into this
energy reserve is that galactic shear
when coupled to the cloud-complex length scale causes these
complexes to rotate with a much higher velocity
than observed.   Furthermore, detailed numerical simulations
show that it is in fact rather difficult to couple the galactic rotation
energy into internal motion energy of the cloud (Das and Jog 1995).

We introduce below a mechanism which can effectively
tap into the energy reserve of galactic rotation without
causing significant vortical motion of the clouds.  
The mechanism operates through the mediation of the spiral density wave,
and is a byproduct of the energy and angular momentum
exchange process between the density wave and the disk matter
at the quasi-steady state of the wave mode.

\section{Energy Injection into the Star-Gas Two-Fluid
through the Spiral Collisionless Shock}

For an open spiral wave mode, the potential and density
spiral patterns are phase-shifted from each other in azimuth.  Inside
corotation, the potential spiral lags the density spiral and
vice versa outside corotation (Zhang 1996).  
The existence of the phase shift 
indicates that there is a secular torque by
the spiral wave on the disk matter, and, at the
quasi-steady state of the wave mode, a secular energy and
angular momentum transfer between the disk matter and the density wave,
which is mediated by a local gravitational instability 
at the spiral arms (Zhang 1996).

Associated with the energy and angular momentum transfer 
between the disk matter and the wave there is heating of 
the disk matter, due to the fact that
the wave pattern speed $\Omega_p$ is in general
not equal to the angular speed of the matter $\Omega$.
Specifically, we have that the rates of
loss of orbital energy and angular
momentum of the basic state matter per unit area
are related through
\begin{equation}
{ {dE_{basic~state}} \over {dt}} = 
\Omega { {dL_{basic~state}} \over {dt}} ,
\end{equation}
and the rates of gain
of energy and angular momentum
by the wave are related through
\begin{equation}
{ {dE_{wave}} \over {dt}} = \Omega_p { {dL_{wave}} \over {dt}}
.
\end{equation}
\noindent Since $dL_{basic~state}/dt$
is equal in magnitude to $dL_{wave} /dt$,
it follows that the
rate of random energy gain per unit area of the disk matter
is related to the angular momentum exchange rate per unit area
through
\begin{equation}
{ {d \Delta E} \over {dt}}
\equiv
{ {d (E_{basic~state} - E_{wave})} \over {dt}} = (\Omega-\Omega_p)
{ {dL_{wave}} \over {dt}}
,
\end{equation}
where $L_{wave}$ is the angular momentum density of the wave.
Note that this expression is true (and has a positive sign) both
inside and outside corotation, since both $\Omega - \Omega_p$ and
$dL_{wave}/dt$ change sign across corotation.

The above expression can be further written in terms of the 
spiral parameters by (Zhang 1998)
\begin{equation}
{ {d \Delta E} \over {dt}}
= {1 \over 2} (\Omega - \Omega_p)
F^2 v_c^2 \tan i \sin(m \phi_0) \Sigma_0
,
\label{eq:e1}
\end{equation}
where $i$ is the pitch angle of the spiral, $m$ is the number of
spiral arms, $\phi_0$ the potential and density phase shift, $F$ the
fractional amplitude of the spiral, $v_c$ the circular speed of
the galaxy, and $\Sigma_0$ the surface density of the disk.  
Equation (\ref{eq:e1}) gives the rate of random energy increase of matter
per unit area, valid for {\em both} the stellar
{\em and} the gaseous components.  For the former, it has been previously
shown that this process could quantitatively account for the age-velocity
dispersion relation of the solar neighborhood stars (Zhang 1999).

\section{The Rate of Energy Injection and Rate of Energy
Cascade}

For the case of the ISM,
we check out below that the above rate of
energy injection (with the injected energy being the orbital
energy converted into random energy by the spiral density
wave) is comparable to the rate of downward turbulent energy cascade.

The average rate of energy injection per unit mass 
into the orbiting disk matter,
using the fitted value of the stellar velocity
diffusion coefficient $D^{3d}$
of D$^{3d}$=6 $\times$ 10$^{-7}$ (km/sec)$^2$ yr$^{-1}$ 
(Wielen 1977; Zhang 1999), is
\begin{equation}
{ {d \Delta E} \over {dt}} = {1 \over 2}  {{ d \Delta v^2 } \over {dt}}
= {1 \over 2} D^{3d} = 3 \times 10^{-7} (km/sec)^2 yr^{-1}
.
\end{equation}

On the other hand, the rate of energy cascade is, 
using the eddy turnover formula of von Weizsacker (1951) and
using the standard values of the Galactic molecular cloud 
velocity disperson of 10 km/sec and complex size scale 1 kpc
\begin{equation}
{{\Delta v^3}
\over {L}}
= { { (10 km/sec)^3} \over { 1000 pc} }
= 8.7 \times 10^{-7} (km/sec)^2 yr^{-1}
.
\end{equation}
The two rates are quite comparable, taking into account of the fact that
the average energy injection rate calculated above is 
an underestimate of the instantaneous energy injection rate
during the crossing of the $\sim$ 1 kpc width spiral arm shock, 
since this average is taken over the time period of the entire 
orbital cycle, which included
the long inter-arm migration period during which there is
no energy injection from the spiral shock.

The analyses of the data for the Carina molecular complex
region (Zhang et al. 2001) showed that the
observed tight size-line-width correlation for this region could
not have been due to the bipolar outflow or the energy injection
from stellar radiation, since the correlation is the same no
matter a given cloud is near or away from the outflow, near
or away from the young stars.  There is also no known supernovae
or superbubble in the surroundings of this region to serve as a
possible energy injection source (Kornreich \& Scalo 2000).
Evidence suggests that
the spiral density wave associated with the Carina arm was responsible
for both a significant part of the cloud-heating of this region 
and for producing the observed size-line-width correlation which holds for
this region of over 150 parsec in extent (Zhang et al. 2001). 

The spiral density wave mechanism is capable of injecting energy 
to the disk matter on all scales, since it is operated through the 
gravitational potential on the disk matter, 
which eliminated the need of a direct coupling
of energy from the size scale of 1 kpc to a few scales downward.  

One interesting question would be the turbulent motion for the gas in
the flocculent galaxies.
Many flocculent galaxies are found to have more coherent underlying spiral
structures in near infrared imaging (Thornley 1997).  Even for those that
do not possess grand design spirals, as long as there is significant 
evolution of galaxy morphology during which gravitational energy is 
constantly being converted into the random motion energy of the particles, 
the turbulence is still being continuously powered.


\begin{thebibliography}{}

\bibitem[\protect\citeauthoryear{Das}{1995}] {}
Das, M. and C.J. Jog.
\newblock {\em ApJ}, 451, 167, 1995

\bibitem[\protect\citeauthoryear{Kornreich \& Scalo}{2000}] {}
Kornreich, P. and J. Scalo.
\newblock {\em ApJ}, 531, 366, 2000

\bibitem[\protect\citeauthoryear{larson}{1981}] {}
Larson, R.B.
\newblock {\em MNRAS}, 186, 479, 1981;
{\em MNRAS}, 194, 809, 1981

\bibitem[\protect\citeauthoryear{Thornley}{1997}] {}
Thornley, M.D.
\newblock {\em Ph.D. Dissertation}, Univ. Maryland, College Park,
1997

\bibitem[\protect\citeauthoryear{Wielen}{1977}] {}
Wielen, R. \newblock {\em A\&A}, 60, 263, 1977

\bibitem[\protect\citeauthoryear{von Weizsacker}{ 1951}] {}
von Weizsacker, C.F. 1951, 
\newblock {\em ApJ}, 114, 165, 1951

\bibitem[\protect\citeauthoryear{Zhang}{1996}] {}Zhang, X.
\newblock {\em ApJ}, 457, 125, 1996; {\em ApJ}
499, 93, 1998; {\em ApJ}, 518, 613, 1999

\bibitem[\protect\citeauthoryear{Zhang et al.}{2001}] {}
Zhang, X., Lee, Y., Bolatto, A. and A.A. Stark.
\newblock {\em ApJ}, 553, 274, 2001

\end{thebibliography}
\end{document}